# BOOSTER CAVITY DAMPER REDESIGN FOR PIP-II

Dustin Pieper†, Brian Vaughn, Robyn Madrak
Fermilab, Batavia, IL, USA

*Abstract*

A new Higher Order Mode (HOM) damper was designed and tested for the Booster accelerator cavity at Fermilab. In anticipation of the PIP-II upgrade, it was discovered that the higher beam intensity of PIP-II may cause beam instability due to an excited mode at 105 MHz. This unfortunately corresponds with the cavity's 2nd order harmonic mode, which will sweep from 86-105.7 MHz.

The new damper is a modification of an existing damper that was designed to reduce an existing static HOM at 83 MHz, with the new design intending to cover the 2nd order HOM as well. The existing damper uses an inductive coupling loop to extract RF energy from the cavity which then goes through a filter in order to reflect the fundamental frequency back into the cavity while passing HOMs to a dump load. The new damper intends to replace the filter portion of the system with a wider band variant while also changing the topology from a coaxial cable loop filter to a componentized PCB-based design.

Primary design challenges include bandwidth coverage, impedance matching of the various modes, long term thermal and mechanical stability, radiation hardness, and high voltage handling. Initial designs achieved the desired damping but were found to quickly succumb to destructive arcing due to the voltages present. More finalized designs intend to address this problem through circuit redesigns and the use of hardier components.

## EXISTING DAMPER SYSTEM

In the Fermilab Booster accelerating cavity, in addition to the fundamental mode used to accelerate protons, there exist a number of unwanted higher order modes (HOMs). These HOMs occur as a result of the cavity's geometry and can be excited by the beam [1]. HOMs are undesirable as they create unwanted acceleration of the beam in ways that create beam instability. To mitigate these HOMs, passive damper structures are added to the cavity to remove energy from the HOMs and dissipate it outside of the cavity.

Dampers are composed of three features, (1) an inductive or capacitive coupler inside the cavity that extracts energy from the cavity at various frequencies, which are then sent through (2) a filter that reflects the fundamental mode back into the cavity (to prevent wasting desired RF accelerating energy) while passing the unwanted HOMs through to (3) a resistive load. Currently there exist three passive dampers for handling HOMs at 83MHz, 167 MHz, and 220 MHz [2]. The 167 MHz and 220 MHz HOMs represent third and fourth order harmonics of the fundamental mode at 53 MHz. For clarification, since the Booster's fundamental mode actually tunes from 38 MHz to 53 MHz over the course of a cycle, with the HOMs also moving by a similar amount as well, the modes are defined here according to their frequency at the top of the cycle. This is also because these modes are at their highest power level at the top of the cycle, meaning damping is most critical at those frequencies. Meanwhile, the 83 MHz mode is a static mode related to features in the cavity unaffected by the tuners and thus doesn't move in frequency over the cycle. A diagram of the 83 MHz damper, as well as pictures of its installed parts is shown below in Figure 1.

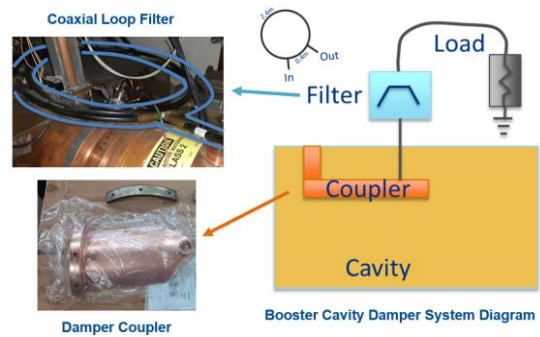

Figure 1: Diagram of 83 MHz damper and pictures of physical parts

Note that for the 83 MHz mode, there are actually two dampers on either side of the cavity, each with their own filter. The coupler for the 83 MHz damper is an inductive loop made from a plate formed to the curvature of the cavity and shorted at one end (which also serves as its physical support in the cavity) while also connected to an external N-type port connector on the other end, from which coupled energy is extracted out of the cavity. A split-path filter is connected to this port in order to only pass the 83 MHz mode. This filter is realized through the use of two lengths of looped coaxial cable that create a path split at a T-junction at the connector port before being re-combined at an output T-junction, which subsequently leads to a high-power load located upstairs in the Booster gallery. The two coaxial path lengths, which are 0.4 m and 2.4 m, respectively, are chosen such that the recombined fundamental signal cancels at the output, forming a standing wave node at that location. This prevents the fundamental mode from being sent to the load while still passing the 83 MHz mode, damping the HOM.

## 105 MHZ MODE DAMPING

After the completion of the PIP-II upgrade of the Fermilab facility, the beam intensity going to Booster will be increased. At this new intensity, there is concern that the 2nd order mode which peaks at 105 MHz will start to be a source of beam instability. Although this mode exists in the cavity at current intensity levels, it is not strong enough to be considered a source of concern but will be after PIP-II. To address this, the filter of the 83 MHz damper is being redesigned in order to also damp the 105 MHz mode while still maintaining its current performance at 83 MHz.

In designing a new filter, it was decided to change the split-path cable design of the current filter to a PCB-based componentized filter design. This change was pursued for a number of reasons. For one, a split-path filter is inherently limited in its design flexibility due to reliance on path length for defining its filtering behaviour. This is especially troublesome for filtering a second-order harmonic while blocking the first-order harmonic, since the first-order harmonic's standing wave nodes all only occur in places where the second-order harmonic has nodes as well, meaning that they can't be filtered separately. By comparison, a PCB based LC filter has significantly more design flexibility and can be easily extended with additional stages. Additionally, a PCB design is more desirable operationally, as the PCB would be far more compact compared to the meter-wide loop of cable used now.

That being said, there were a number of challenges to overcome in designing the new filter. For one, the cavity impedance looking into the damper's coupler port is not a 50 Ω system but instead has a complex impedance. Furthermore, the impedance varies significantly with frequency, and it can be difficult to pin down the impedance value for a given resonance mode. Since most RF simulation tools expect a single input impedance, this can make it difficult to simulate circuit behaviour over the bandwidth of operation. In order to facilitate the design process, the system impedance was instead back calculated using the known behavior of the existing damper, iterating on the value until the simulated behaviour matched. The notional model impedance used was eventually determined to be $20+j*35\Omega$. While this static value doesn't capture the frequency variability of the impedance, and also likely is less valid at 105 MHz, it acted as a useful baseline for design iteration purposes and was found to provide replicatable results in later testing.

From this starting point, the Advanced Design System (ADS) simulator tool was used to iterate through candidate circuit designs while assuming the baseline impedance. When a functional design was found, the S-parameter data could then be extracted from the ADS simulation and applied to an Ansys HFSS full-wave simulation of the cavity. This full-wave simulation provided a more realistic impedance profile of the cavity by simulating the physical behaviours of the cavity. Note that to use the S-parameter port definition tools in HFSS, the ADS S-parameter values had to be converted to be referenced to 50Ω again. Once a design had been verified, it could proceed to the physical prototype stage. From this process, the following design was conceived and built, as presented in Figure 2 and Figure 3.

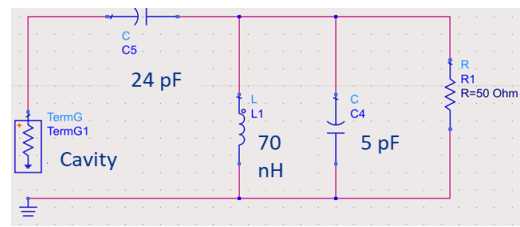

Figure 2: Schematic of prototype damper filter.

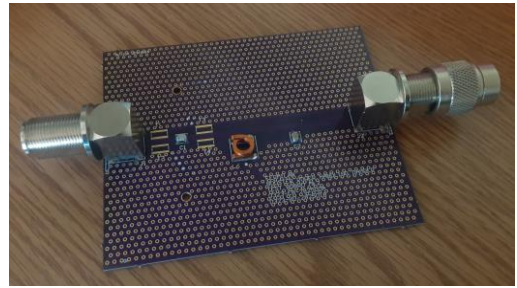

Figure 3: Constructed prototype of damper filter.

Key features of this design include high voltage components, extra solder pads near the input capacitor for adding thermal bridges if necessary, and via flooding for heat dissipation. Additionally, the board was mounted inside a ventilated enclosure with fan for heat removal.

With two of these filters built for the two 83 MHz dampers, testing on an actual cavity could begin. The dampers were added to the cavity's damper ports, and the cavity was tuned to its fundamental frequency of 53 MHz. This test was conducted at low power, with the cavity being excited by a VNA connected to a coupling loop added onto the cavity's power amplifier (PA). An S21 measurement was then made on the cavity between the PA coupling loop and the cavity's gap monitor in order to measure resonant behaviour. The resonant peaks at each mode could then be compared with similar measurements done with the old damper filter and with no filter in order to determine damping effectiveness. The results of these measurements are given below in Figure 4.

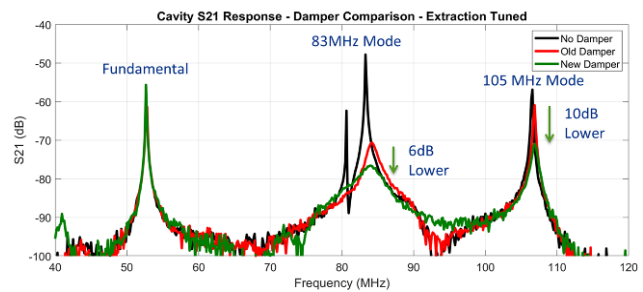

Figure 4: Measured cavity S21 for different damper filters

Overall, it was found that the new design provided about 10 dB more attenuation than the original filter, while also improving on the filtering of the 83 MHz mode by 6 dB. Given this result, high power testing commenced next to ensure that the filter can survive the voltages expected from the cavity.

## HIGH POWER TESTING

High power testing of the damper filter is intended to verify two important concerns ahead of long-term installation. The first is to ensure that the filter can withstand the high voltages that might be expected on the cavity's damper port. The second is to ensure that the filter can survive the heat generated by the power moving through it. Although the filter only has reactive components, these still have small amounts of resistance, and so any current driven through them will create heat which can reduce component lifetime or even cause failure.

High power testing is similar to low power testing in that it involves installing the damper filters on the Booster cavity. However, for high power testing, the cavity is driven directly by the power amplifier, as it would be in the accelerator. Ideally, the PA would be driven with a program similar to the Booster accelerator ramp program in order to emulate real life behaviour, such as regular pulsing and the subsequent heating and cooling cycles. However, upon conducting the test, it was found that the circuit failed immediately upon application of power due to an over-voltage condition at the connector. The aftermath of this failure is show below in Figure 5.

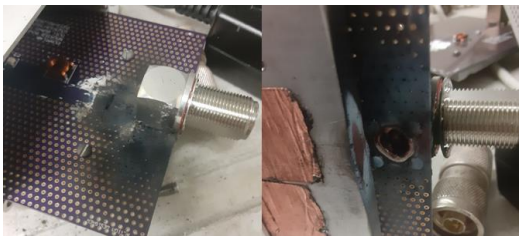

Figure 5. Damage caused by arcing on new damper filters

*Path Forward*

Since that test, updates to the design have been explored to try and reduce the voltages at the port connector, as well as further improve design robustness. Analysis of the failed circuit suggested that the input capacitor acted as an open circuit to the fundamental mode, which doubles the voltage seen at the input. The goal of the new design is to trade reflected voltage for current (representing a rotation of the Smith chart away from the open-circuit condition). The new design can be seen below in Figure 7 and 7.

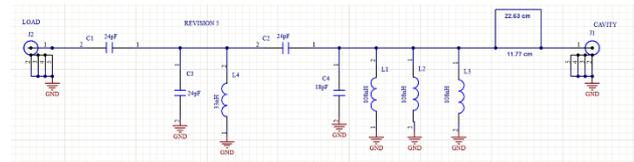

Figure 6. Schematic for new damper prototype

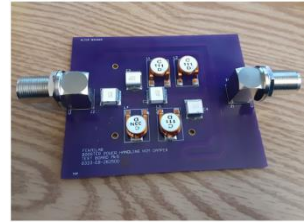

Figure 7. New prototype for improved voltage handling

The new design removes the input capacitor and instead adds a short split-path section on the front of the circuit in order to mitigate voltage at the connector, which is the weak point of the design. Another potential design avenue would be to replace the N-type connectors with DIN connectors which have much higher voltage handling capability, although this would require replacing the damper connector on the cavity as well. While this new design has passed low power testing, it is still awaiting high power testing which is contingent on certain facility improvements being completed.

## CONCLUSION

New damper filters for the Fermilab Booster cavity passive damper system are being designed. These new filters will allow the existing dampers to also damp a mode at 105 MHz which may otherwise cause instability at future beam intensities. This new design is being implemented as a PCB mounted LC circuit, and initial testing has verified its ability to filter the 105 MHz mode while still damping the modes of its predecessor. However, the design has had issues with high voltage testing which are still being addressed, with testing of a new iteration set to begin soon.